\documentclass[preprint,superscriptaddress,eqsecnum,nofootinbib]{revtex4}
\usepackage{graphicx,amssymb,amsbsy,bm,amsmath,psfrag,dsfont}
\usepackage{slashed}
\usepackage{hyperref} \usepackage[dvips]{color}

\begin{document}

\newcommand{\be}{\begin{equation}}
\newcommand{\ee}{\end{equation}}
\newcommand{\bea}{\begin{eqnarray}}
\newcommand{\eea}{\end{eqnarray}}
\newcommand{\no}{\noindent}

\newcommand{\la}{\lambda}
\newcommand{\si}{\sigma}
\newcommand{\vk}{\vec{k}}
\newcommand{\vx}{\vec{x}}
\newcommand{\om}{\omega}
\newcommand{\Om}{\Omega}
\newcommand{\ga}{\gamma}
\newcommand{\Ga}{\Gamma}
\newcommand{\gaa}{\Gamma_a}
\newcommand{\al}{\alpha}
\newcommand{\ep}{\epsilon}
\newcommand{\app}{\approx}
\newcommand{\uvk}{\widehat{\bf{k}}}
\newcommand{\OM}{\overline{M}}

\title{Spacetime Emergence and General Covariance Transmutation}
\author{Chiu Man Ho} \email{chiuman.ho@vanderbilt.edu}
  \affiliation{Department of
  Physics and Astronomy, Vanderbilt University, Nashville, TN 37235, USA}
\author{Thomas W. Kephart}\email{tom.kephart@gmail.com}
\affiliation{Department of
  Physics and Astronomy, Vanderbilt University, Nashville, TN 37235, USA}
\author{Djordje Minic}\email{dminic@vt.edu}
\affiliation{Department of Physics, Virginia Tech, Blacksburg, VA 24061, U.S.A.}
\author{Y. Jack Ng}\email{yjng@physics.unc.edu}
\affiliation{Institute of Field Physics, Department of Physics and Astronomy,
University of North Carolina, Chapel Hill, NC 27599, U.S.A.}

\date{\today}

\begin{abstract}

Spacetime emergence refers to the notion that classical spacetime ``emerges" as
an approximate macroscopic entity from a non-spatio-temporal structure present
in a more complete theory of interacting fundamental constituents. In this
article, we propose a novel mechanism involving the ``soldering" of internal
and external spaces for the emergence of spacetime and the twin transmutation
of general covariance. In the context of string theory, this mechanism points
to a critical four dimensional spacetime background.


\end{abstract}

\maketitle


\section{Emerging Theories on Emergence}

Is spacetime a necessary feature of a fundamental description of nature?  This
problem has long been debated among physicists and philosophers; and in recent
years, it has been
exacerbated by the unsolved nature of
quantum gravity.  For one thing, a straightforward quantization of general
relativity leads to uncontrollable divergences.  The non-renormalizability of
gravity has naturally led to the suggestion that general relativity
is perhaps only a low energy effective theory,
and accordingly the metric and connection forms are nothing more than the
collective or hydrodynamic fields of some fundamental
degrees of freedom.  Then, it is a logical possibility that
the physical spacetime described by Einstein's theory of
gravity is actually emergent
--- a feature akin to hydrodynamics emerging from molecular dynamics or
nuclear physics from quantum chromodynamics.

The idea that spacetime is not a fundamental physical concept and that
it emerges from some other more fundamental notions is not new.
For example, Sakharov argued more than forty years ago that
gravity is an induced force, and the dynamical spacetime is emergent
from some more fundamental ``atomic'' structure \cite{sakharov,visser}.
Adler et al. \cite{adler} explored an alternative approach
to a microscopic theory of gravitation,
in which the gravitational fields are identified with photon pairing
amplitudes of a superconducting type.

In the 1970s, the remarkable discoveries regarding black hole
thermodynamics
by Bekenstein, Hawking and others \cite{bhth,Hawking} have pointed in a more precise
fashion towards an emergent
nature of gravity and thus the emergent nature of spacetime \cite{emergent1,emergent2,emergent3}.
If indeed gravitation can be understood as a thermodynamic effect, the statistical
origin of such a course grained notion is immediately called into question.
More recently, these intuitions from semiclassical quantum gravity have been
reexamined by using the notion of the holographic principle \cite{holo1,holo2}
in the context of the celebrated AdS/CFT correspondence in string theory
\cite{adscft}. More generally, various string dualities suggest that spacetime is
indeed only a derived concept \cite{seiberg}.  It has also been argued
\cite{branva} that string gas cosmology favors four emergent
macroscopic spacetime dimensions.

Emergent spacetime is also suggested in some other theory candidates
of quantum gravity.  The equations of
loop quantum gravity \cite{smolin,rovelli}
do not presuppose spacetime; instead they are expected to give rise to
spacetime at distances large compared to the Planck length.
In the causal dynamical triangulation program \cite{AJL},
a non-perturbative sum over geometries has yielded a fully dynamical
emergence of a classical
background (and solution to the Einstein equations) with the correct
dimensionality of four
at large scales (and effectively two at the Planck scale).

Last but not least, the concept of emergence, which is so natural in the
context of many-body
physics, has been argued to apply more generally in the context of emergence
of all fundamental forces in physics \cite{wen,volovik,hu}.
Obviously, the idea of emergent spacetime
has solid roots from many points of view.

In this article, we point out a new and explicit mechanism for the
emergence of spacetime and general covariance.  Then, we argue that,
in the context of string theory, such a mechanism
points to a critical four dimensional spacetime background.

\section{Emergent Cosmological Constant}

Consider a D-dimensional spacetime with diagonal metric
$G_{\mu \nu}$,
where $\mu,\nu=0,1,...,D-1$. On this spacetime, suppose
there exists a D-component non-linear sigma field $\phi^{a}$.
The diagonal internal-space metric, which determines
the dynamics of the non-linear sigma fields, is given by the metric
$h_{ab}$, where $a, b=0,1,...,D-1$.
Notice that we have chosen the number of non-linear sigma fields
to be exactly the same
as the dimension of the spacetime. Thus, throughout the entire article,
both of the spacetime and internal-space indices
run from $0$ to $D-1$.

Let us consider a theory with the D-dimensional Einstein gravity coupled to
the D-dimensional non-linear sigma model, with action given by
\bea
\label{action}
S=\int\,d^{D}x\, \sqrt{G} \left\{\,-\frac12\, M^{D-2} \,R +\frac{1}{\la^2}\,
G^{\mu\nu}\,h_{ab}(\phi)\,\partial_{\mu}\,\phi^a\,\partial_\nu\,\phi^b\,\right\}\,,
\eea
where $R$ is the Ricci scalar of the D-dimensional spacetime,
$1/\lambda^2$ is a coupling constant, and $M$ is the D-dimensional Planck mass.

Now, suppose we impose the following ansatz:
\bea
\label{ident}
\phi^{0} &=& i\,\alpha \, t + C\,, \\
\label{ident2}
\phi^{a} &=& \alpha \, x^{a} + C\,, ~~~~~\textrm{for}~~ a\neq 0\,,
\eea
where $\alpha$ and $C$ are constants carrying mass dimensions of
$\frac{D-2}{2}+1$ and $\frac{D-2}{2}$ respectively.
Notice that the extra factor of $i$ in the identification
\eqref{ident} is required to properly match
the Lorentzian and Euclidean signatures of the D-dimensional
spacetime and internal-space respectively.
In fact, this is the same ansatz that was imposed in
\cite{Omero:1980vx,GellMann:1984sj,GellMann:1985if,Chakrabarty:1989uf} in an
attempt to induce dynamical compactification
and hence inflation, although exponential inflation failed to emerge.
A successful model using this ansatz to generate exponential inflation
has recently been provided in \cite{Ho:2010vv}. Interestingly,
't Hooft made a similar ansatz
in \cite{Hooft} in his consideration of Higgs mechanism for gravity.
Physically, this ansatz is
equivalent to identifying
the components of the non-linear sigma fields with the spacetime
coordinates, which is intuitively reasonable
because the non-linear sigma fields themselves are ``coordinates"
of the internal-space manifold. Mathematically, the ansatz is
also justified as the non-linear sigma fields are functions of the
spacetime coordinates, and it simply specifies an explicit
dependence of the non-linear sigma fields on the spacetime coordinates.

One may wonder if there is any dynamical basis that leads to
the ansatz \eqref{ident} and \eqref{ident2}. We will not address this issue here;
it is certainly
beyond the scope of the present paper.
In fact, even the papers by Gell-Mann and Zwiebach
\cite{GellMann:1984sj,GellMann:1985if} and 't Hooft \cite{Hooft} have not provided
any dynamical basis for the ansatz. In what follows, we
will just adopt this ansatz to see what it implies. But, as we will show in
Section III, it is
precisely the form of the ansatz shown in
\eqref{ident} and \eqref{ident2} that is required for
general covariance. Hence the results will provide an {\it a posteriori} justification
of the ansatz choice.

With this ansatz, the action $S$ now becomes
\bea
\label{action2}
S
= \int\,d^{D}x\, \sqrt{G} \,\left\{\,-\frac12\, M^{D-2} \,R + \frac{1}{\la^2}\,
\alpha^2 \; \left(\,-G^{00}\,h_{00}+G^{ij}\,h_{ij}\,\right)
\,\right\}\,.
\eea

Actually, the identifications \eqref{ident} and \eqref{ident2} should not
affect the tensor character of
$h_{ab}$. So we expect the components of $h_{\mu\nu}$ to be proportional
to the components of $G_{\mu\nu}$.
For a given D-dimensional spacetime, if we assume that the
internal-space metric is such that
\bea
\label{hG}
h_{00} = \beta \, G_{00} \, ~~~~~\textrm{and}~~~~
h_{ij} = -\beta \, G_{ij},
\eea
with $\beta >0$ being a constant,
then the action can be written as
\bea
\label{action3}
S = \int\,d^{D}x\, \sqrt{G} \,\left\{\,-\frac12\, M^{D-2} \,R - \frac{1}{\la^2}\,
\alpha^2\,\beta \; G^{\mu\nu}\,G_{\mu\nu}
\,\right\}\,.
\eea
Since $\frac{1}{D} \,G^{\mu \nu} \, G_{\mu \nu} =1$, it follows that
\bea
\label{action4}
S &=& -\frac12\, M^{D-2} \;\int\,d^{D}x\, \sqrt{G} \,\left(\;R + 2 \,
\frac{1}{\la^2} \,\frac{D\,\alpha^2\,
\beta }{M^{D-2}} \;\,\right)\,.
\eea
Thus, a cosmological constant of magnitude $\frac{1}{\la^2} \,
\frac{D\,\alpha^2\, \beta }{M^{D-2}}$ emerges. Reversing the process, we can absorb the
cosmological constant in a non-linear sigma field and gain conceptual
advantage as we will discuss below.

As an illustrative and realistic example of the above idea, we consider a 4-dimensional spacetime with
the Friedmann-Robertson-Walker (FRW) metric in Appendix A.

\section{General Covariance Transmutation}

Since both sets of indices $\{\mu,\nu\}$ and $\{a,b\}$ run from 0 to $D-1$,
we can form the following
interactions terms and add them to the action in \eqref{action}:
\bea
 \textrm{(I)}&& ~~~~ f\;G_{\mu \nu} \, h^{ab} \, \partial_a\, \phi^\mu\, \partial_b\, \phi^\nu \\
 \textrm{(II)}&& ~~~~ g\; G^{\mu \nu} \, h_{\mu \nu} \, (\partial_a \,\phi^{a})^2
\eea
Obviously, both  (I) and (II) break general convariance,
because spacetime and internal-space indices have been
contracted in a mixed way.
However,  general convariance can \emph{re-emerge} as follows.
If we impose the identifications \eqref{ident} and \eqref{ident2},
the non-linear sigma fields essentially become the spacetime coordinates.
There is no longer a difference between internal-space indices
and spacetime indices. Instead,
the spacetime indices coincide exactly with the internal-space indices.
Originally, $h_{ab}(\phi)$ is a function of the non-linear sigma fields.
But after making the identifications
\eqref{ident} and \eqref{ident2},\, $h_{ab}$
becomes an explicit function of spacetime coordinates.
This means that when we write $h_{ab}$, the indices $a$ and
$b$ become spacetime indices. In this sense, all of the spacetime
or internal-space indices in
$f\;G_{\mu \nu} \, h^{ab} \, \partial_a\, \phi^\mu\, \partial_b\,
\phi^\nu$ and $g\;G^{\mu \nu} \, h_{\mu \nu} \, (\partial_a \,\phi^{a})^2$
have been contracted.

In light of the above identifications \eqref{ident} and \eqref{ident2},
the action becomes
\vspace{0.2cm}
\bea
\label{actionCovariant}
S'
= \int\,d^{D}x\, \sqrt{G} \,\left\{\,-\frac12\, M^{D-2} \,R +\left(\,\frac{1}{\la^2} + f \, \right)\,\alpha^2
\; \left(\,-G^{00}\,h_{00}+G^{ij}\,h_{ij}\,\right)
\,\right\}\,,
\eea
where we are forced to set $g=0$ because the term
$g\;G^{\mu \nu} \, h_{\mu \nu} \, (\partial_a \,\phi^{a})^2$ leads to
a factor of \,$(D-1+i)^2\,\alpha^2$,\, rendering the action non-Hermitian and hence non-unitary.
One could argue that this term may still be unitary if it is PT-symmetric \cite{Bender}, although we will not
consider this possibility here as it is not crucial to our discussions.

Since there is now no more uncontracted spacetime indices in $S'$,
we conclude that general covariance has
emerged upon the identifications \eqref{ident} and \eqref{ident2}.
As a result, the action $S'$ is generally covariant.

We can understand the origin of the emergent general covariance
as follows.
Originally, we have two different sets of diffeomorphisms. One
is in the internal space and the other is the non-compact spacetime.
Let's call their diffeomorphism algebras $D_1$ and $D_2$
respectively. Before the two sets of diffeomorphisms
are coupled  we have $D_1 \otimes D_2$
as the symmetry algebra of the theory. Once they are coupled,
only the diagonal subalgebra
$D \subset (\,D_1 \otimes D_2\,)$ is preserved. The purpose of adopting
the ansatz \eqref{ident} and \eqref{ident2} is precisely
to extract the digaonal subalgebra, which leads to the general covariance in \eqref{actionCovariant}.
Moreover, we note that the idea of ``soldering" internal and external indices has been
used by Polyakov in the context of non-critical string theory \cite{polyakov}, where it is argued
that general covariance emerges from the underlying (gauge) current algebra structure.

Similar to the case in the previous section,
if we assume \eqref{hG},
then the action $S'$ can be written as
\bea
\label{action5}
S' &=& -\frac12\, M^{D-2} \;\int\,d^{D}x\, \sqrt{G} \,
\left\{\;R + 2 \,\left(\,\frac{1}{\la^2}+f\,\right) \,\frac{D\,\alpha^2\,
\beta }{M^{D-2}} \;\,\right\}\,.
\eea
As a result, a shifted (with respect to eq.(\ref{action4}))
cosmological constant of magnitude\\
 $\left(\,\frac{1}{\la^2}+f\,\right) \,\frac{D\,\alpha^2\,
\beta }{M^{D-2}}$ emerges.

Therefore, we have the following conclusions. First of all,
even if $f=0$, a cosmological constant can
already emerge upon the identifications \eqref{ident}
and \eqref{ident2} . When $f\neq 0$, the same identifications \eqref{ident}
and \eqref{ident2} lead to another contribution to the cosmological
constant. In addition, the introduction of the term
$f\;G_{\mu \nu} \, h^{ab} \, \partial_a\, \phi^\mu\, \partial_b\, \phi^\nu$
breaks general covariance.
However, with the identifications \eqref{ident} and \eqref{ident2},
the broken general covariance (which exists in the first place)
can be restored. This ``emergence" of general covariance
can be called ``general covariance
transmutation".  General covariance is a gauge symmetry and is the
starting point of general relativity.  With the transmutation of
general covariance at hand, the stage is set for the emergence of
spacetime itself.  But for the latter, we need another theoretical input as we will see in Section IV.

Finally, we would like to provide an analysis of the equation of motion for the non-linear sigma fields in order to make sure that everything is consistent with the anzatz invoked in \eqref{ident} and \eqref{ident2}. Since the term $g\;G^{\mu \nu} \, h_{\mu \nu} \, (\partial_a \,\phi^{a})^2$ leads to a non-Hermitian factor of \,$(D-1+i)^2\,\alpha^2$\, when the ansatz is applied, we will only consider the term $f\;G_{\mu \nu} \, h^{ab} \, \partial_a\, \phi^\mu\, \partial_b\, \phi^\nu$ in addition to the action in \eqref{action}. The equation of motion for a fixed component field $\phi^\beta$ is then given by
\bea
\label{EOM}
&& 2\,\partial_\mu \left(\,G^{\mu\nu}\, h_{\beta a}\,\partial_\nu\,\phi^a\,\right) + 2\,f\,\lambda^2\, \partial_a \left(\,G_{\beta\nu}\, h^{ab}\,\partial_b\,\phi^\nu\,\right)
\nonumber \\
&& - \frac{\partial}{\partial\,\phi^\beta}\,\left(\,G^{\mu\nu}\,h_{ab}\,\partial_{\mu}\,\phi^a\,\partial_\nu\,\phi^b \,\right)
- f\, \lambda^2\, \frac{\partial}{\partial\,\phi^\beta}\,\left(\,G_{\mu \nu} \, h^{ab} \, \partial_a\, \phi^\mu\, \partial_b\, \phi^\nu \,\right) =0\,.
\eea
One can easily verify that the equation of motion is satisfied by the ansatz in \eqref{ident} and \eqref{ident2} only if
$h_{00} \propto G_{00}$\, and \,$h_{ij} \propto G_{ij}$ which are precisely what we imposed in \eqref{hG}.


\section{Emergent Space(-time)}

Let us now apply the mechanism of ``soldering" internal and external spaces to string theory.  Imagine that we start
with 2-dimensional (2d) gravity with a cosmological constant. This can be
viewed as the sigma model action for 2d string theory
(in what will turn out to be a 2d spacetime background, so $c,d=0,1$)
\bea
\label{actionstr}
S_2=\frac{T}{2}\,\int\,d^{2}\sigma\, \sqrt{g} \;\left[\;
g^{\rho\tau}\,\xi_{cd}(X)\,\partial_{\rho}\,X^c\,\partial_\tau\,X^d\,+\cdots\; \right]\,,
\eea
where $T$ is the string tension, $g_{\rho\tau}$ is the worldsheet metric, $\xi_{cd}(X)$ is the metric of the target manifold and
``$\cdots$" represents the remaining terms in the full 2d string action \cite{polchinski}.
This starting 2d
sigma model is crucial on three scores.  Firstly it is renormalizable, secondly
the 2d gravity is merely topological, and thirdly the $\beta$-function
of the background metric viewed as a coupling gives, to leading order in the inverse of the
string tension,
the Einstein equations of motion (reflecting the famous fact of
perturbative string theory \cite{string1,string2})
\bea
L\, \frac{d\, \xi_{cd} (X)}{ d\,L} = R_{cd}\,,
\eea
where $L$ is the renormalization group scale of the sigma model and $R_{cd}$ is the Ricci tensor.
The 2d string can also be given a non-pertrubative formulation in terms of the $c=1$ matrix model \cite{moore},
which represents a regularization of the Polyakov path integral.

Now, we can envision increasing the string coupling until
the worldsheet description is
replaced by something more fundamental. One indication, from M-theory,
is that the string ``thickens'' into a membrane \cite{duff},
described by the corresponding 3d sigma (membrane) model
(where now it will turn out that $c,d = 0,1,2$)
\bea
\label{actionmem}
S_3=\frac{T}{2}\,\int\,d^{3}\sigma\, \sqrt{g} \;\left[\;
g^{\rho\tau}\,\xi_{cd}(X)\,\partial_{\rho}\,X^c\,\partial_\tau\,X^d \,+\cdots\;\right]\,,
\eea
where $g_{\rho\tau}$ becomes the metric of the worldvolume and ``$\cdots$" represents the remaining terms in the
full 3d membrane action \cite{duff}. Note that this 3d
sigma model is non-renormalizable and it cannot be taken as
the short-distance definition of the theory. A suitable definition is supplied by
a regularization of the 3d sigma model action known as
Matrix theory \cite{bfss,banks}.

As we saw in Section III, the mechanism of ``soldering" internal and external indices could
lead to the twin transmutation of general covariance.
By this mechanism, the worldvolume general covariance could
induce the target-space
general covariance and thus the ambient target spacetime.
Finally, by invoking the holographic principle,
one can propose that
the 3d membrane theory (coupled to some other degrees of freedom)
is a holographic
boundary dual of some emergent bulk 4d spacetime description.
The radial direction in the 4d spacetime emerges without
being a space dimension in the 3d
field theory.  It can be interpreted as the renormalizable group scale, or
the energy scale used to probe the 3d theory.

In short, the emergence of spacetime comes about when we go from weak to
strong coupling (strings to membranes) and when we reinterpret the
renormalization group flow of this membrane theory coupled to some other
matter holographically as a 4d theory that involves 4d gravity.
All these insights are supported by what we know from
string theory \cite{string1,string2}.
It is pleasing that the
mechanism of ``soldering" internal and external indices discussed in this article not only points to
emergent spacetime but, in the context of string theory, also
to the critical 4d spacetime background.

We have just witnessed space emergence twice (with spatial dimensions grown by two).
It is now natural to ask whether time can also emerge.
Note that the current knowledge of string theory relies on the definition of the
underlying sigma model in terms of a {\it Euclidean} quantum (conformal) field theory.
The Lorentzian background is obtained after the usual Wick rotation, which
in the case of a general time-dependent background (and dynamical
causal structure), is problematic. One might say that holography \cite{adscft} resolves the
issues associated with dynamical causal structure. But holography relies
on the existence of a large spacetime and the associated asymptotic regions which simply do not exist
in general. Given what we know about string theory both in the perturbative and holographic contexts,
our proposal should be understood in the same vein.

Of course, the unity of
space and time demanded by relativity would seem to imply that time also emerges, given that space
emerges.  On the other hand, given the manifold issues associated with time-dependence in
string theory, it has been repeatedly argued that the emergence of time
may violate locality and conceivably also causality \cite{seiberg}.
This would imply that
time is fundamental even if space is emergent \cite{djetal1,djetal2,djetal3}.  Furthermore, it seems unlikely
that any physical system can evolve without an underlying time.  But this is
not an air-tight argument against emergent time --- even the dearth of examples of which
may merely be due to our lack of imagination.
In any case, the problem of time emergence has profound
implications for the physics of black hole and
cosmological singularities, and may well shed light on the structure of
the Hilbert space of the Universe and the origin of the
arrow of time \cite{seiberg,djetal1,djetal2,djetal3,carroll}.  However, it is quite possible that
time emergence is even harder to fathom than space emergence, and that
the true physics of emergent time would require even more radical ideas \cite{djetal1,djetal2,djetal3,chia1,chia2,chia3,chia4}
than the ones presented in this paper.



\vskip 1cm

\noindent
{\bf Acknowledgments:}
CMH and TWK were supported by the US Department of Energy
by grant no. DE-FG05-85ER40226, DM by
DE-FG05-92ER40677 and YJN by DE-FG02-06ER41418.

\appendix

\section{~An Example Involving the Friedmann-Robertson-Walker Metric}

Consider a 4-dimensional spacetime with the Friedmann-Robertson-Walker (FRW) metric:
\bea
\label{metric}
ds^2 = dt^2- a^2(t) \,g_{i j} \,dx^{i} \,dx^{j}\,,
\eea
where $i,j=1,2,3$\, and $a(t)$ is the scale factor. The various components of the Ricci tensor $R_{\mu\nu}$
resulting from the above metric (\ref{metric}) are
\bea
\label{R0}
R_{00} &=& -3\,\frac{\ddot{a}(t)}{a(t)}\,G_{00}\,, \\
\label{Ri}
R_{ij} &=& - \left(\, \dot{H}(t)+ 3 \, H^2(t)\,\right)\,G_{ij} \,,
\eea
where $G_{00}=1$,\, $G_{ij}=-a^2(t)\, g_{ij}$\, and\, $H(t)=\frac{\dot{a}(t)}{a(t)}$.

On the other hand, the equations of motion following from the
variation of the action $S$ in \eqref{action2} with respect to the metric
$G^{\mu\nu}$ are given by
\bea
\label{R00}
R_{00}&=& - \frac{4\,\alpha^2}{M^{2}}\,\frac{1}{\la^2} \, h_{00} \,,\\
\label{Rij}
R_{ij}&=&\frac{4\,\alpha^2}{M^{2}}\,\frac{1}{\la^2} \, h_{ij} \,.
\eea

By Einstein's field equation, the components \eqref{R00} and \eqref{Rij} represent the energy source that leads to
the geometry realized by the metric \eqref{metric}. Thus, we proceed to solve the equations of motion
\eqref{R00} and \eqref{Rij} by using the geometry
of the metric \eqref{metric}. This requires matching \eqref{R00} with \eqref{R0} and \eqref{Rij} with \eqref{Ri}.
Hence, the metric of the internal-space manifold is required to satisfy:
\bea
h_{00} &=&  3\,\frac{\ddot{a}(t)}{a(t)} \, \frac{M^2}{4 \,(\frac{1}{\la^2})\,\alpha^2}\,G_{00}\,, \\
h_{ij} &=&  - \left(\, \dot{H}(t)+ 3 \, H^2(t)\,\right)\,\frac{M^2}{4\, (\frac{1}{\la^2})\,\alpha^2}\,G_{ij}\,.
\eea

In particular, if \,$3\,\frac{\ddot{a}(t)}{a(t)} = \dot{H}(t)+ 3 \, H^2(t)$= constant, the internal-space metric is of the
form $h_{00} = \beta \, G_{00} $ and $h_{ij} = -\beta \, G_{ij} $. In this case, the only solution allowed is
$a(t)\propto e^{h\,t}$ with $h$ being a constant. As shown above, the emergent cosmological constant is given by
$\Lambda=\frac{4}{\la^2} \,\frac{\alpha^2\,\beta }{M^{2}}$, for $D=4$. This gives
the conventional result, namely
$h=\sqrt{\frac{\Lambda}{3}}$, when the universe is dominated by the cosmological constant.

\end{document}